
%

\magnification=\magstep1	          
\raggedbottom

\parskip=9pt

\def\singlespace{\baselineskip=12pt}      
\def\sesquispace{\baselineskip=14pt}      
\def\Ch{K}








\font\openface=msbm10 at10pt
 %

\def\Minkowski     {{\hbox{\openface M}}}

 %
 %
 %






\def\card{\mathop{\rm card}\nolimits}    


\def\past   {\mathop {\rm past     }\nolimits}
\def\future {\mathop {\rm future   }\nolimits}


%

%
%



\def\implies{\Rightarrow}

%


 \def\dal{\displaystyle{{\hbox to 0pt{$\sqcup$\hss}}\sqcap}}



\def\lto{\mathop
        {\hbox{${\lower3.8pt\hbox{$<$}}\atop{\raise0.2pt\hbox{$\sim$}}$}}}
\def\gto{\mathop
        {\hbox{${\lower3.8pt\hbox{$>$}}\atop{\raise0.2pt\hbox{$\sim$}}$}}}
%
%
%



\def\&{{\phantom a}}

\def\braces#1{ \{ #1 \} }



\def\to{\mathop\rightarrow}	

\def\ideq{\equiv}		

\def\SetOf#1#2{\left\{ #1  \,|\, #2 \right\} }


\def\union{\cup}
\def\intersect{\cap}


\def\interior #1 {  \buildrel\circ\over  #1}     




\def\basisvector#1#2#3{
 \lower6pt\hbox{
  ${\buildrel{\displaystyle #1}\over{\scriptscriptstyle(#2)}}$}^#3}

\def\eps{\epsilon}





%


%
%
%
%
%
%
%
%
%
%

%
 \let\miguu=\footnote
 \def\footnote#1#2{{$\,$\parindent=9pt\baselineskip=13pt%
 \miguu{#1}{#2\vskip -7truept}}}
%
%

\def\linebreak{\hfil\break}


\def\BulletItem #1 {\item{$\bullet$}{#1 }}

\def\ReferencesBegin
{
 \singlespace					   
 \vskip 0.5truein
 \centerline           {\bf References}
 \par\nobreak
 \medskip
 \noindent
 \parindent=2pt
 \parskip=6pt			
 }

\def\section #1 {\bigskip\noindent{\bf #1 }\par\nobreak\smallskip\noindent}

\def\subsection #1 {\medskip\noindent[ {\it #1} ]\par\nobreak\smallskip}

\def\reference{\hangindent=1pc\hangafter=1} 

\def\ref{\reference}

\def\journaldata#1#2#3#4{{\it #1 } {\bf #2:} #3 (#4)}
 %

\def\eprint#1{$\langle$#1\hbox{$\rangle$}}
 %


\def\author#1 {\medskip\centerline{\it #1}\smallskip}

\def\address#1{\centerline{\it #1}\smallskip}

\def\furtheraddress#1{\centerline{\it and}\smallskip\centerline{\it #1}\smallskip}

\font\titlefont=cmb10 scaled\magstep2 


\font\csmc=cmcsc10  

\def\Remark{\noindent {\csmc Remark \ }}
\def\Conjecture{\noindent {\csmc Conjecture \ }}





\def\QuotationBegins
{
 \singlespace                                
 \smallskip\leftskip=1.5truecm\rightskip=1.5truecm     
 \noindent	
 } 
\def\QuotationEnds
{
 \smallskip\leftskip=0truecm\rightskip=0truecm       
 \sesquispace
 \par				
 \noindent
 } 


\def\refbreak{\baselineskip=15pt\hfil\break\singlespace}

\def\interval{\hbox{interval}}
\def\Prob{\hbox{Prob}}

\font\csmc=cmcsc10  

\def\Remark{\noindent {\csmc Remark \ }}
\def\Conjecture{\noindent {\csmc Conjecture \ }}

%


\rightline{gr-qc/0309009}
\rightline{SU--GP--03/01--02}

\sesquispace
\centerline {\titlefont Causal Sets: Discrete Gravity} 
\bigskip
\centerline {\it Notes for the Valdivia Summer School, Jan. 2002}

\bigskip


\singlespace			        

\author{Rafael D. Sorkin}

\address
 {Department of Physics, Syracuse University, Syracuse, NY 13244-1130, U.S.A.}
\furtheraddress
 {Queen Mary College, University of London, Mile End Road, London E1 4NS}

\smallskip

\centerline {\it \qquad\qquad internet address: sorkin@physics.syr.edu}

\smallskip



\noindent
{\it 
These are some notes in lieu of the lectures I was scheduled to give,
but had to cancel at the last moment.  In some places, they are more
complete, in others much less so, regrettably.  I hope they at least
give a feel for the subject and convey some of the excitement felt at
the moment by those of us working on it.}

\noindent
{\it 
An extensive set of references and a glossary of terms can be found
at the end of the notes.
For a philosophically oriented discussion of some of the
background to the causal set idea, see reference
[1].  For general background see [2]
[3] [4] [5] [6]
[7].}


\sesquispace

\bigskip\medskip


\section{Introduction}
It seems fair to say that causal set theory has reached a stage
in which questions of phenomenology are beginning to be
addressed meaningfully.  This welcome development is due on one
hand to improved astronomical observations which shed light on
the magnitude of the cosmological constant (in apparent
confirmation of a long-standing prediction of the theory)
and on the other hand to theoretical advances which for the
first time have placed on the agenda the development of a
quantum dynamical law for causal sets (and also for a scalar
field residing on a background causal set).  What we have so far
are: $(i)$ an apparently confirmed order of magnitude prediction
for the cosmological constant; $(ii)$ a method of counting black
hole horizon ``states'' at the kinematical level; $(iii)$ 
the beginnings of a
framework in which two-dimensional Hawking radiation can be
addressed; $(iv)$ a {\it classical} causal set dynamics which
arguably is the most general consistent with the discrete
analogs of general covariance and relativistic causality; and in
consequence of this, both $(v)$ the formulation of a ``cosmic
renormalization group'' which indicates how one might in
principle solve some of the large number puzzles of
cosmology without recourse to a post-quantum era of ``inflation''; 
and
$(vi)$ a hint of how non-gravitational matter might arise
at the fundamental level from causal sets rather than having to
be added in by hand or derived at a higher level {\`a} la
Kaluza-Klein from an effective spacetime topology arising from
the fundamental structures via coarse-graining.  In addition, a
good deal of computer code has been written for use in causal
set simulations, including a library of over 5000 lines of Lisp
code that can be used by anyone with access to the Emacs editor.
At present, the principal need, in addition to fleshing out the
developments already outlined, is for a quantum analog of the
classical dynamics alluded to above.
It looks as if a suitable quantum version of Bell causality (see below)
could lead directly to such a dynamics, that is to say, to a theory of
quantum spacetime, and in particular to a theory of quantum gravity.

The remainder of these notes rapidly reviews the subject in its current
state, progressing broadly from kinematics to dynamics to
phenomenology.  Although this sequence does not always reflect exactly
the chronological development of the theory, it is not far off, and it
also fits in well with Taketani's ``3-stages'' schema of scientific
discovery [8].

\section{Origins of the causet idea}
The tradition of seeing the causal order of spacetime as its most
fundamental structure is almost as old as the idea of spacetime itself
(in its Relativistic form).  In [9], Robb presented a set of
axioms for Minkowski space analogous to Euclid's axioms for plane
geometry.  In so doing, he effectively demonstrated that, up to an
overall conformal factor, the geometry of 4-dimensional flat spacetime
(which I'll denote by $\Minkowski4$, taken always with a definite
time-orientation) can be recovered from nothing more than the underlying
point set and the order relation $\prec$ among points (where $x \prec y
\iff$ the vector from $x$ to $y$ is timelike or lightlike and
future-pointing).  Later, Reichenbach [10] from the side of
philosophy and Zeeman [11] from the side of mathematics
emphasized the same fact, the latter in particular by proving the
theorem, implicit in [9], that any order-isomorphism of
$\Minkowski4$ onto itself must --- up to an overall scaling --- belong
to the (isochronous) Poincar{\'e} group.

In a certain sense, however, these results appear to say more than they
really do.  Informally, they seem to tell us that $\Minkowski4$ can be
reconstructed from the relation $\prec$, but 
in actually carrying out the reconstruction (see below), 
one needs to know that what one 
is trying to recover is a flat spacetime and not just a conformally
flat one.  Clearly, there's nothing in the relation $\prec$ {\it per se}
which can 
tell us that.  This difficulty shows itself, in a sense, in the failure
of Zeeman's theorem for $\Minkowski2$ and $\Minkowski1$ (i.e. 1+1 and
0+1 dimensional Minkowski space).  But it shows up still more clearly
with the curved spacetimes of General Relativity, where the natural
generalization of the flat space theorems is that a Lorentzian geometry
$M$ can be recovered from its causal order only up to a {\it local}
conformal factor.

Notice that when one says that a Lorentzian manifold $M$ is recovered,
one is talking about {\it all} the mathematical structures that go into
the definition of a spacetime geometry: its topology, its
differential structure and its metric.  Various special results show how
to recover, say, the topology (see e.g. [12]) but the most
complete theorems are those of [13] and [14], the latter
delineating very precisely how close $M$ can come to violating causality
without the theorem breaking down.

The upshot of all these reconstruction theorems 
is that, in the continuum, something is
lacking if we possess only the causal order, namely the {\it conformal
factor} or equivalently the ``volume element'' $\sqrt{-g}\,d^4x$.  On
the other hand, if we do give the volume element (say in the form of a
measure $\mu$ on $M$) then it is clear that metric $g_{ab}$ will be
determined in full.  The causal order alone, however, is --- in the
continuum --- incapable of furnishing such a measure.

This failing is perhaps one reason to question the reality of the
continuum, but of course it is not the only one.
In modern times, doubts show up clearly in Riemann's inaugural lecture
(Habilitationsschrift) [15], where he contrasts the idea of what he calls a
{\it discrete manifold} with that of a continuous manifold, which latter
he takes to be a relatively unfamiliar and unintuitive idea in
comparison with the former!  The most evocative quotes from this lecture
are perhaps the following:


\QuotationBegins
  Gr\"ossenbegriffe sind nur da m\"oglich, wo sich ein
  allgemeiner Begriff vor\-findet, der verschiedene Bestimmungsweisen
  zul\"asst.  Je nachdem unter diesen Bestimmungsweisen von einer zu einer
  andern ein stetiger {\"U}bergang stattfindet oder nicht, bilden sie eine
  stetige oder discrete Mannigfaltigkeit; die enzelnen Bestimmungsweisen
  heissen im ersten Falle Punkte, im letzten Elemente dieser
  Mannigfaltigkeit. (p.273)
\QuotationEnds
or in translation,\footnote{$^\star$}
{These translations are not guaranteed, but they're a lot better than what
  ``Google'' did (try it for fun!).}
\QuotationBegins
   Concepts of magnitude are only possible where 
   a general concept is met with 
   that admits of different individual instances [Bestimmungsweisen]. 
   According as, among these individual instances, a continuous passage 
   from one to another takes place or not, they form a continuous or
   discrete manifold; the individual instances are called in the first
   case points, in the second elements of the manifold;
\QuotationEnds
\QuotationBegins
    Die Frage {\"u}ber die G{\"u}ltigkeit der Voraussetzungen der Geometrie im
    Unendlichkleinen h{\"a}ngt zusammen mit der Frage nach dem innern Grunde
    der Massverh{\"a}ltnisse des Raumes.  Bei dieser Frage, welche wohl
    noch zur Lehre vom Raume gerechnet werden darf, kommt die obige
    Bemerkung zur Anwendung, dass bei einer discreten Mannigfaltigkeit
    das Princip der Massverh{\"a}ltnisse schon in dem Begriffe dieser
    Mannigfaltigkeit enthalten ist, bei einer stetigen aber anders woher
    hinzukommen muss.  Es muss also entweder das dem Raume zu Grunde
    liegende Wirkliche eine discrete Mannigfaltigkeit bilden, oder der
    Grund der Massverh{\"a}ltnisse ausserhalb, in darauf wirkenden bindenden
    Kr{\"a}ften, gesucht werden.  
\QuotationEnds
or in translation,
\QuotationBegins
    The question of the validity of the presuppositions of geometry in
    the infinitely small hangs together with the question of the inner
    ground of the metric relationships of space. [I almost wrote
    ``spacetime''!] In connection with the latter question, which
    probably [?] can still be reckoned to be part of the
    science of space, the above remark applies, that for a discrete
    manifold, the principle of its metric relationships is already
    contained in the concept of the manifold itself, whereas for a
    continuous manifold, it must come from somewhere else.
    Therefore, either the reality which underlies physical space must
    form a discrete manifold or else the basis of its metric
    relationships must be sought for outside it, in binding forces
    [bindenden Kr{\"a}fte] that act on it;
\QuotationEnds
and finally,
\QuotationBegins
   Bestimmte, durch ein Merkmal oder eine Grenze unterschiedene Theile
   einer Mannigfaltigkeit heissen Quanta.  Ihre Vergleichung der
   Quantit{\"a}t  nach geschieht bei den discreten Gr{\"o}ssen durch
   Z{\"a}hlung, bei den stetigen durch Messung. (p.274)
\QuotationEnds
or in translation,
\QuotationBegins
   Definite portions of a manifold, distinguished by a criterion
   [Merkmal] or a boundary, are called quanta.  Their quantitative
   comparison happens for discrete magnitudes through counting, for
   continuous ones through measurement.
\QuotationEnds

With the subsequent development of physics, more compelling reasons
emerged for questioning the continuum, including the singularities and
infinities of General Relativity, of Quantum Field Theory  (including the
standard model), and of black hole thermodynamics.
Einstein, for example, voiced doubts of this sort very early 
[16]:

\QuotationBegins
But you have correctly grasped the drawback that the continuum brings.
If the molecular view of matter is the correct (appropriate) one, i.e., if
a part of the universe is to be represented by a finite number of moving
points, then the continuum of the present theory contains too great a
manifold of possibilities.  I also believe that this too great is
responsible for the fact that our present means of description miscarry
with the quantum theory.  The problem seems to me how one can formulate
statements about a discontinuum without calling upon a continuum
(space-time) as an aid; the latter should be banned from the theory as a
supplementary construction not justified by the essence of the problem,
which corresponds to nothing ``real''.  But we still lack the mathematical
structure unfortunately.  How much have I already plagued myself in this
way!
\QuotationEnds
and at a later stage stressed the importance of the causal order in
this connection, writing [17] that it would be
``especially difficult to derive something like a spatio-temporal
quasi-order'' from a purely algebraic or combinatorial scheme.


The causal set  idea is, in essence, nothing more than an attempt to combine
the twin ideas of discreteness and order to produce a structure on which
a theory of quantum gravity can be based.  That such a step was almost
inevitable is indicated by the fact that very similar formulations were
put forward independently in [4], [5] and [2], after
having been adumbrated in [18].  
The insight underlying these
proposals is that, in passing from the continuous to the discrete, one
actually {\it gains} certain information, because ``volume'' can now be
assessed (as Riemann said) {\it by counting\/}; and with both order {\it
and} volume information present, we have enough to recover geometry.

In this way the topology, the differential structure and, the metric of
continuum physics all become unified with the causal order (much as mass
is unified with energy in Special Relativity).  Moreover the Lorentzian
signature (namely $(- + + +)$ in 4 dimensions) is singled out as the only
one compatible with a consistent distinction between past and future,
hence the only one that can make contact with the idea of causal order.


To see how these basic ideas work themselves out, we need first a more
precise statement of what a causal set is.

\section{What is a causal set?}
As a mathematical structure, a causal set (or {\it causet} for short) is
simply a {\it locally finite ordered set}.  In other words, it is a set
$C$ endowed with a binary relation $\prec$ possessing the following
three properties:
\item{} $(i)$ transitivity: 
  $(\forall x,y,z\in C)(x\prec y\prec z\implies x\prec z)$
\item{} $(ii)$ irreflexivity:
  $(\forall x\in C)(x\not\prec x)$
\item{} $(iii)$ local finiteness:
  $(\forall x,z\in C) \, ( \card \SetOf{y\in C}{x\prec y\prec z} < \infty )$

\noindent
where `$\card$' stands for ``cardinality''.In the presence of
transitivity, irreflexivity automatically implies 
acyclicity, i.e. the absence of cycles
$x_0\prec{x_1}\prec{x_2}\prec\cdots\prec x_n=x_0$, and this is often
taken as an axiom in place of $(ii)$.  The condition $(iii)$ of local
finiteness is a formal way of saying that a causet is {\it discrete}.
Thus the real number line, for example does not qualify as a causet,
although it is a partial order.\footnote{$^\dagger$}
{The above definition utilizes the so called ``irreflexive convention''
 that no element precedes itself.  Axioms $(i)$ and $(ii)$ define what
 is variously called an ``order'', a ``partial order'', a ``poset'', an
 ``ordered set'' or an ``acyclic transitive digraph''.  Axiom $(iii)$,
 which expresses the condition of local finiteness, can also be stated
 in the form ``every order-interval has finite cardinality''.}

A structure satisfying the above axioms can be thought of as a
graph, and in this sense is conveniently represented as a
so-called Hasse diagram in which the elements of $C$ appear as
vertices and the relations appear as edges.  (The sense of the
relation is usually shown, just as in the spacetime diagrams of
Relativity theory, by making the line between $x$ and $y$ be a
rising one when $x\prec{y}$.)  Actually, it is not necessary to
draw in all the relations, but only those not implied by
transitivity (the ``links''), and this convention is almost
always adopted to simplify the appearance of the diagram.  A
causet can also be thought of as a matrix $M$ (the ``causal
matrix'') with the rows and columns labeled by the elements of
$C$ and with the matrix element $M_{jk}$ being $1$ if $j\prec k$
and $0$ otherwise.
Perhaps, however, the most suggestive way to think of a causet
for the purposes of quantum gravity is as a relation of
``descent'', effectively a family tree that indicates which of
the elements $C$ are ``ancestors'' of which others.

The multiplicity of imagery associated with causets (or partial orders
more generally) is part of the richness of the subject and makes it
natural to use a variety of language in discussing the structural
relationships induced by the basic order relation $\prec$.  Thus, the
relationship $x\prec y$ itself, is variously described by saying that
$x$ {\it precedes} $y$, that $x$ is an {\it ancestor} of $y$, that $y$
is a $descendant$ of $x$, or that $x$ lies to the {\it past of} $y$ (or
$y$ to the {\it future} of $x$).  Similarly, if $x$ is an {\it
immediate} ancestor of $y$ (meaning that there exists no intervening $z$
such that $x\prec z \prec y$) then one says that $x$ is a {\it parent}
of $y$, or $y$ a {\it child} of $x$, or that $y$ {\it covers} $x$, or
that $x\prec y$ is a {\it link}.  (See the glossary.)

Still other interpretations of the relation $\prec$ are possible
and can also be useful.  For example a causet of finite
cardinality is equivalent to a $T_0$ topological space of finite
cardinality, allowing one to use the language of topology
in talking about causets (which indeed may turn out to have
more than just a metaphorical significance).  A causet can also
be treated as a {\it function} by identifying $C$ with the
function `past' that associates to each $x\in C$ the set
$\past(x)$ of all its ancestors, and this is in fact the
representation on which the Lisp code of [19] is
based.

For the purposes of quantum gravity, a causal set is, of course,
meant to be the deep structure of spacetime.  Or to say this
another way, the basic hypothesis is that spacetime ceases to
exist on sufficiently small scales and is superseded by an
ordered discrete structure to which the continuum is only a
coarse-grained, macroscopic approximation.

Now, at first sight, a structure based purely on the concept of
order might seem to be too impoverished to reproduce the
geometrical and topological attributes in terms of which general
relativistic spacetime is normally conceived.  However, if one
reflects that light cones can be defined in causal terms and
that (in the continuum) the light cones determine the metric up
to a conformal rescaling, then it becomes understandable that
(given minimal regularity conditions like the absence of closed
causal curves) the causal order of a Lorentzian manifold (say
$J^+$ in the usual notation) captures fully the conformal
metric, as well as the topology and the differential structure.
The volume element $\sqrt{-g}d^nx$ cannot be recovered from $J^+$,
but in the context of a {\it discrete} order, it can be obtained
in another way --- by equating the {\it number} of causet
elements to the {\it volume} of the corresponding region of the
spacetime continuum that approximates $C$.
As discussed above, these observations provide the
kinematical starting point for a theory of discrete quantum
gravity based on causal sets.  The dynamics must then be
obtained in the form of a ``quantum law of motion'' for the
causet.  Let us consider the kinematics further.



\section{Causal set kinematics in general}
Both the study of the mathematics of causets for its own sake
and its study for the sake of clarifying how the geometrical and
topological properties of a continuous spacetime translate into
order properties of the underlying causet can be regarded as
aspects of causal set  kinematics: the study of causets without
reference to any particular dynamical law.

A large amount is known about causet kinematics as a
result of extensive work by both physicists and mathematicians.
(See for example [20] [21]
[22] [23] [24] [25]
[26] [27].)  
(Some of the mathematicians were directly
influenced by the causal set idea, others were studying ordered
sets for their own reasons.)  We know for example, that the
length of the longest chain\footnote{$^\flat$}
{The term `chain' is defined in the glossary.}
provides a good measure of the proper time (geodesic length)
between any two causally related elements of a causet that can
be approximated by a region of Minkowski space [28].
And for such a causet, we also possess at least two or three
good dimension estimators, one of which is well understood
analytically [29].

The next few sections are devoted to some of these topics.

\section{``How big'' is a causet element?}
Of course the question is badly worded, because a causet element has no
size as such.  What it's really asking for is the conversion factor 
$v_0$ for which $N=V/v_0$.  Only if we measure length in units such that
$v_0\ideq1$
can we express the hypothesis that number=volume in the form $N=V$.  On
dimensional grounds, one naturally expects $v_0\sim(G\hbar)2$ [where
I've taken $c\ideq 1$].  But we can do better than just relying on
dimensional analysis {\it per se\/}.  Consider first the entropy of a
black hole horizon, which is given by $S=A/4G\hbar=2\pi{A}/\kappa$,
with $\kappa=8\pi{}G$, the rationalized gravitational constant. 
This formula suggests forcefully that about one bit of entropy belongs
to each horizon ``plaquette'' of size $\kappa\hbar$, and that the
effectively finite size of these ``plaquettes'' reflects directly an
underlying spacetime discreteness.  Consideration of the so called
entanglement entropy leads to the same conclusion, namely that there
exists an effective ``ultraviolet cutoff'' at around
$l=\sqrt{\kappa\hbar}$. [30]

A related but less direct train of thought starts by considering the
gravitational action-integral ${1\over 2\kappa}\int RdV$.  Here the
``coupling constant'' $1/2\kappa$ is an inverse length$^2$ and
conventional Renormalization Group wisdom suggests that, barring any
``fine tuning'', the order of magnitude of such a coupling constant will
be set by the underlying ``lattice spacing'', or in this case, the
fundamental discreteness scale, leading to the same conclusion as before
that $l\ideq(v_0)^{1/4}$ is around $l=\sqrt{\kappa\hbar}\sim 10^{-32}cm$.


A noteworthy implication of the formula $l\sim\sqrt{\kappa\hbar}$ is
that $l\to0$ if $\hbar\to0$ ($\kappa$ being fixed).  That is, the
classical limit is necessarily a continuum limit: spacetime discreteness
is inherently quantal.

\section{The reconstruction of $\Minkowski4$}
In order to get a better feel for how it is that ``geometry = order +
number'', it is useful to work through the reconstruction --- in the
continuum --- of $\Minkowski4$ from its causal order and
volume-element.
%
%
The proof can be given in a quite constructive form which I'll only
sketch here.

We start with a copy $M$ of $\Minkowski4$ and let $\prec$ be its causal
order.  We construct in turn: light rays $l$, null 3-planes, spacelike
2-planes, spacelike lines, arbitrary 2-planes, arbitrary lines, parallel
lines, parallelograms, vectors.  Once we have vector addition (affine
structure) it is easy to get quadratic forms and in particular the flat
metric $\eta_{ab}$.  (The normalization of $\eta_{ab}$ uses the
volume information.)

You may enjoy working these constructions out for yourself, so I won't
give them all here.  At the risk of spoiling your fun however, let me
give just the first two, which perhaps are less straightforward than the
rest.  We define a light ray $l$ to be a maximal chain such that 
$\forall x,y\in l, \interval(x,y)$ is {\it also} a chain; and from any
such $l$ we get then the null hyperplane $N(l)=l\union l^{\natural}$, 
where $l^\natural$ is the set of all points of $M$ spacelike to $l$.


\section{Sprinkling, coarse-graining, and the ``Hauptvermutung''}
A basic tenet of causet theory is that spacetime does not exist at the
most fundamental level, that it is an ``emergent'' concept which is
relevant only to the extent that some manifold-with-Lorentzian-metric
$M$ furnishes a good approximation to the physical causet $C$.  Under
what circumstances would we want to say that this occurred?  So far the
most promising answer to this question is based on the concepts of {\it
sprinkling} and {\it coarse-graining}. 

Given a manifold $M$ with Lorentzian metric $g_{ab}$ (which is, say,
globally hyperbolic) we can obtain a causal set 
$C(M)$ by selecting points of $M$ and endowing them with the order
induced from that of $M$ (where in $M$, $x\prec{y}$ iff there is a
future causal curve from $x$ to $y$).  In order to realize the equality
$N=V$, the selected points must be distributed {\it with unit density}
in $M$.  One way
%
%
to accomplish this (and conjecturally the {\it only} way!) is to
generate the points of $C(M)$ by a {\it Poisson process}.  (To realize
the latter, imagine dividing $M$ up into small regions of volume $\eps$
and independently putting a point into each region with probability
$\eps$.  In the limit $\eps\to0$ this is the Poisson process of unit
intensity in $M$.)  Let us write $M\approx C$ for the assertion that $M$
is a good approximation to $C$.  The idea then is that $M\approx C$ if
$C$ ``might have been produced by a sprinkling of $M$'' (in a sense to
be specified more fully below).

It's important here that the elements of $C(M)$ are selected at
random from $M$.  In particular, this fact is an ingredient in the
heuristic reasoning leading to the prediction of a fluctuating
cosmological constant (see below).  But such a {\it kinematic}
randomness might seem gratuitous.  Wouldn't a suitable {\it regular}
embedding of points into $M$ yield a subset that was equally uniformly
distributed, if not more so?  To see what goes wrong, consider the
``diamond lattice'' in $\Minkowski2$ consisting of all points with
integer values of the null coordinates $u=t-x$ and $v=t+x$.  This would
{\it seem} to be a uniform lattice, but under a boost $u\to\lambda u$,
$v\to v/ \lambda$ it goes into a distribution that looks entirely
different, with a very high density of points along the $u$=constant
lines (say) and large empty spaces in between.  In particular, our
diamond lattice is far from Lorentz invariant, which a truly uniform
distribution {\it should be} --- and which $C(M)$ produced by a Poisson
process {\it actually is}.  Examples like this suggest strongly that, in
contrast to the situation for Euclidean signature, only a random
sprinkling can be uniform for Lorentzian signature. 


I have just argued that the idea of random sprinkling must play a role
in making the correspondence between the causet and the continuum, and
for purely kinematic reasons.  A second concept which might be needed as
well, depending on how the dynamics works out, is that of {\it
coarse-graining}.  Indeed, one might expect that, on very small scales,
the causet representing our universe will no more look like a manifold
than the trajectory of a point particle looks microscopically
like a smooth curve in
nonrelativistic quantum mechanics.  Rather, we might recover a manifold
only after some degree of ``averaging'' or ``coarse-graining'' 
(assuming also that we keep away from the big bang and from black hole
interiors, etc., where we don't expect a manifold at all).
That is, we might expect not $C\approx{M}$ but $C'\approx{M'}$, where 
$C'$ is some coarse-graining of $C$ and $M'$ is $M$ with a
correspondingly rescaled metric.  
The relevant notion of coarse-graining here seems to be an analog of
sprinkling applied to $C$ itself:  
let $C'$ be obtained from $C$ by selecting a subset at random, keeping each
element $x\in{C}$ with some fixed probability, say $1/2$ if we want a
2:1 coarse-graining.  


Implicit in the idea of a manifold approximating a causet is that the
former is relatively unique; for if two very different manifolds could
approximate the same $C$, we'd have no objective way to understand why
we observe one particular spacetime and not some very different one.
(On the other hand, considering things like AdS/CFT duality, who
knows...!)  
The conjecture that such ambiguities don't occur has been called the
``Hauptvermutung''.  In the $G\hbar\to0$ limit, it has already been
proven in [31].  Moreover, the fact that we know how
to obtain dimensional and proper time information in many situations
(see below) is strong circumstantial evidence for its truth at finite
$G\hbar$.  Nevertheless it would be good to prove it in full,
in something like the following form.

\Conjecture  If $M_1\approx C$ and $M_2\approx C$ then $M_1\approx M_2$

\noindent
Here $M_1\approx M_2$ means that the manifolds $M_1$ and $M_2$ are
``approximately isometric''.  As the quotation marks indicate, it is
surprisingly difficult to give this conjecture rigorous meaning, due
ultimately to the Lorentzian signature of $g_{ab}$ (cf. [32]).
Here's a sketch of
how it might be done:  interpret $M\approx C$ to mean that
$\Prob(C(M)=C)$ is relatively large (in comparison with 
$\Prob(C(M)=C')$) for most of the other $C'$; and interpret 
$M_1\approx M_2$ to mean that the random variables  $C(M_1)$ and
$C(M_2)$ share similar probability distributions (on the space of
causets).  These definitions almost make the conjecture look tautological,
but it isn't! (cf. [33])



\section{Dimension and length}

Assuming that --- as seems very likely --- causal sets do possess a
structure rich enough to give us back a macroscopically smooth
Lorentzian geometry, it is important to figure out how in practice one
can extract geometrical information from an order relation.  But before
we can speak of a geometry we must have a manifold, and the most basic
aspect of a manifold's topology is its dimension.  So an obvious first
question is whether there is a good way to recognize the effective
continuum dimension of a causal set (or more precisely of a causal set
that is sufficiently ``manifold like'' for the notion of its dimension
to be meaningful).  In fact several workable approaches exist.  Here are
three of them. 
All three estimators will assign a dimension to an interval $I$ in a causet
$C$ and are designed for the case where $I\approx A$ for some interval
(``double light cone'') $A$ in Minkowski space $\Minkowski^d$.

\noindent {\it Myrheim-Meyer dimension} [5] [29].
Let $N=|I|$ be the number of elements in $I$ 
and let $R$ be the number of relations in $I$
(i.e. pairs $x$, $y$ such that $x\prec y$).  
Let $f(d) = {3\over 2}{3d/2 \choose d}^{-1}$.
Then $f^{-1}(R/{N\choose2})$ is a good estimate of $d$ when 
$N\gg(27/16)^d$.  

\Remark The Myrheim-Meyer estimator is coarse-graining invariant on average (as is
the next)

\noindent {\it Midpoint scaling dimension}.  
Let $I=\interval(a,b)$ and let $m\in I$ be the (or a) ``midpoint'' defined to
maximize $N'=\min\braces{|\interval(a,m)|, |\interval(m,b)|}$.  Then
$\log_2(N/N')$ estimates $d$.

\noindent {\it A third dimension estimator}.  
Let $\Ch$ be the total number of chains in $I$.  Then $\ln{N}/\ln\ln{\Ch}$
estimates $d$.  However, the logarithms mean that good accuracy sets in
only for exponentially large $N$.

\section{A length estimator}
Again this is for $C\approx\Minkowski^d$ (or some convex subspace of
$\Minkowski^d$).  Let $x\prec y$.
The 
most obvious way to define a distance (or better a time-lapse) from
$x$ to $y$ is just to count the number of elements $L$ in the longest chain
joining them, where a joining chain is by definition  a succession
of elements $z_i$ such that $x\prec z_1\prec z_2\prec z_3\cdots\prec y$.  
Clearly a maximal path in this sense is analogous to a timelike
geodesic, which maximizes the proper-time between its endpoints.
It is known that this estimator $L$ converges rapidly to a multiple of
the true proper time $T$ as the latter becomes large.  However the
coefficient of proportionality depends on the dimension $d$ and is known
exactly only for $d=1$ and $d=2$.  For $d>2$ only bounds are known, but
they are rather tight.  


Thus, it seems that we have workable tools for recovering information on
both dimensionality and length (in the sense of timelike geodesic distance).
However, these tools have been proven so far primarily in a flat
context, and it remains to be shown that they continue to work well in
the presence of generic curvature. 


\section{Dynamics}
A priori, one can imagine
at least two routes to a ``quantum causet dynamics''.  On one
hand, one could try to mimic the formulation of other
theories by seeking a causet-invariant analogous to the scalar
curvature action, and then attempting to build from it some
discrete version of a gravitational ``sum-over-histories''.  On
the other hand, one could try to identify certain general
principles or rules powerful enough to lead, more or less
uniquely, to a family of dynamical laws sufficiently
constrained that one could then pick out those members of the
family that reproduced the Einstein equations in an appropriate
limit or approximation.  (By way of analogy, one could imagine
arriving at general relativity either by seeking a spin-2 analog
of Poisson's equation or by seeking the most general field
equations compatible with general covariance and locality.)
%

The recent progress in dynamics has come from the second type of
approach, and with the causet's ``time-evolution'' conceived as a
process of what  may be termed {\it sequential growth\/}.  
That is, the causet is conceived of as ``developing
in time'',\footnote{$^\star$}
{It might be more accurate to say that the growth of the causet
 {\it is time}.}
rather than as ``existing timelessly'' in the manner of a film
strip.  At the same time the growth process is taken to be
random rather than deterministic --- classically random to start
with, but ultimately random in the quantum sense familiar from
atomic physics and quantum field theory.  (Thus the quantum
dynamical law is being viewed as more analogous to a classical
stochastic process like Brownian motion than to a classical
deterministic dynamics like that of the harmonic oscillator.
[34] [35])
Expressed more technically, the idea is to seek a quantum causet
dynamics by first formulating the causet's growth as a classical
stochastic process and then generalizing the formulation to the
case of a ``quantum measure'' [36] or ``decoherence
functional'' [37].


The growth process in question can be viewed as a sequence of ``births''
of new causet elements and each such birth is a transition from one
partial causet to another.  A dynamics or ``law of growth'' is then
simply an assignment of probabilities to each possible sequence of
transitions.  Without further restriction, however, there would be a
virtually limitless set of possibilities for these probabilities.  The
two principles that have allowed us to narrow this field of
possibilities down to a (hopefully) manageable number are {\it discrete
general covariance} and {\it Bell causality}.  To understand the first of
these, notice that taking the births to be sequential implicitly
introduces a labeling of the causet elements (the first-born element
being labeled 0, the second-born labeled 1, etc).  Discrete general
covariance is simply the requirement that this labeling be ``pure
gauge'', that it drop out of the final probabilities in the same way
that the choice of coordinates drops out of the equations of general
relativity.  This requirement has the important side effect of rendering
the growth process Markovian, so that it is fully definable in
terms of {\it transition probabilities} obeying the Markov sum rule.  The
requirement of Bell causality is slightly harder to explain, but it is
meant to capture the intuition that a birth taking place in one region
of the causet cannot be influenced by other births that occur in regions
spacelike to the first region.

Taken together, these assumptions lead to a set of equations and
inequalities that --- remarkably --- can be solved explicitly and in
general [38] [39].  
The resulting probability for a transition
$C\to{}C'$ in which the new element is born with $\varpi$
ancestors and $m$ parents (immediate ancestors) is given by the
ratio
$$
   { \lambda(\varpi,m) \over \lambda(n,0) }  \ ,  \eqno(1)
$$ 
where $n=\card(C)$ is the number of
elements before the birth in question and where 
the function $\lambda$ is defined by the formula
$$
   \lambda(\varpi,m) = \sum_{k=m}^\varpi {\varpi-m\choose k-m} t_k
   \eqno(2)
$$
with $t_n{\ge}0$ and $t_0>0$.
A particular dynamical law, then, is determined by the sequence of
``coupling constants'' $t_n$ (or more precisely, by their ratios).



(It turns out that the probabilities resulting from these rules can be
re-expressed in terms of an ``Ising model'' whose spins reside on the
relations $x{\prec}y$ of the causet, and whose ``vertex weights'' are
governed directly by the parameters $t_n$ [38]. In this way, a
certain form of ``Ising matter'' emerges indirectly from the dynamical
law, albeit its dynamics is rather trivial if one confines oneself to a
fixed background causet.  This illustrates how one might hope to recover
in an appropriate limit, not only spacetime and gravity, but also
certain forms of non-gravitational matter (here unified with gravity in
a way reminiscent of earlier proposals for ``induced gravity'' [40]).)

With some progress in hand concerning both the kinematics and dynamics
of causets, it is possible to start to think about applications, or if
you will ``phenomenology''.  Several projects of this nature are
under way, with some interesting results already obtained.  In the
remaining sections I will mention some of these results and projects.

\section{Fluctuations in the cosmological constant}
From the most basic notions of causal set theory, there follows
already an order of magnitude prediction for the value of the
cosmological constant $\Lambda$.  More precisely, one can argue
that $\Lambda$ should {\it fluctuate} about its ``target
value'', with the magnitude of the fluctuations decreasing with
time in proportion to $N^{-1/2}$, where $N$ is the relevant
number of ancestors (causet elements) at a given cosmological
epoch.  If one assumes that (for reasons yet to be understood)
the target value for $\Lambda$ is zero, and if one takes for the
$N$ of today the spacetime volume of the currently visible
universe from the big bang until the present, then one deduces
for $\Lambda$ a magnitude consistent with the most recent
observations, as predicted already in [41] (and with refined
arguments in [35] and [42]).

Four basic features of causet theory enter as ingredients into the
refined version of the argument: 
the fundamental discreteness,
the relation $n=V$,
the Poisson fluctuations in $n$ associated with sprinkling,
and the fact that $n$ serves as a parameter time in the dynamics of
sequential growth.

From the first of these we derive a finite value of $n$ (at any given
cosmic time).  From the fourth we deduce that, since time is not summed
over in the path-integral of non-relativistic quantum mechanics, neither
should one expect to sum over $n$ in the gravitational path integral that
one expects to result as an approximation to the still to be formulated
quantum dynamics of causets.  But holding $n$ fixed means holding
spacetime volume $V$ fixed, a procedure that leads in the continuum to
what is called ``unimodular gravity''.  

In the classical limit, this unimodular procedure leads 
to the action principle
$$
  \delta\left( \int ({1\over 2\kappa}R - \Lambda_0) dV - \lambda V \right)=0
$$
where 
$\Lambda_0$ is the ``bare'' cosmological constant, 
$V=\int dV$,
$\kappa=8\pi G$,
and
$\lambda$ is a Lagrange multiplier implementing the fixation of $V$.
Plainly, the last two terms combine into $-\int\Lambda dV$ where 
$\Lambda=\Lambda_0+\lambda$, turning the effective cosmological constant 
$\Lambda$ into a free constant of integration rather than a fixed
microscopic parameter of the theory.  Moreover, the fact that $\Lambda$
and $V$ enter into the action-integral in the combination 
$-\Lambda V$ means that they are conjugate in the quantum mechanical
sense, leading to the indeterminacy relation 
$$
               \delta \Lambda \, \delta V \sim \hbar \ .
$$
Finally the Poisson fluctuations in $n$ of size $\delta{n}\sim\sqrt{n}$
at fixed $V$ imply that, at fixed $n$, there will be fluctuations in $V$
of the same magnitude: $\delta{V}\sim\sqrt{n}\sim\sqrt{V}$, which
correspond to fluctuations in $\Lambda$ of magnitude 
$\delta\Lambda\sim\hbar/\delta{V}\sim 1/\sqrt{V}$ (taking $\hbar=1$).
The observed $\Lambda$ would thus be a sort of residual quantum gravity
effect, even though one normally associates the quantum with the very
small, rather than the very big!



Of course, this prediction of a fluctuating $\Lambda$ 
remains at a heuristic level 
until it can be grounded in a complete ``quantum causet dynamics''.  
Nevertheless, given its initial
success, it seems worthwhile to try to extend it by constructing
a model in which not only the instantaneous magnitude of the
fluctuations could be predicted, but also their correlations
between one time and another.  In this way, one could assess
whether the original prediction was consistent with important
cosmological constraints such as the extent of structure
formation and the abundances of the light nuclei.
In
this respect, it is worth noting that current fits of
nucleosynthesis models to the observed abundances favor a
non-integer number of light neutrinos falling between two and
three.  If this indication holds up, it will require some form
of effective negative energy density at nucleosynthesis time,
and a negative fluctuation in the contemporaneous $\Lambda$ is
perhaps the simplest way to realize such an effective density.

\noindent {\it Added note:} A concrete model of the sort suggested above
has been developed in [43].

\section{Links across the horizon}
An important question on which one can hope to shed light while
still remaining at the level of kinematics is that of
identifying the ``horizon states'' that underpin the entropy of
a black hole.  Indeed, just as the entropy of a box of gas is,
to a first approximation, merely counting the molecules in the
box, one might anticipate that the entropy of a black hole is
effectively counting suitably defined ``molecules'' of its horizon.
With this possibility in mind, one can ask whether any simply
definable sub-structures of the causets associated with a given
geometry could serve as candidates for such ``horizon molecules'' in
the sense that counting them would approximately measure the
``information content'' of the black hole.

Perhaps the most obvious candidates of this sort are the causal
links crossing the horizon in the neighborhood of the
hypersurface $\Sigma$ for which the entropy is sought.  (Recall
that a link is an irreducible relation of the causet.)  Of
course, the counting of any small scale substructures of the
causet is prone to produce a result proportional to the area of
the horizon, but there is no reason {\it a priori} why the
coefficient of proportionality could not be divergent or
vanishing, or why, if it is finite, it could not depend on the
details of the horizon geometry.


Djamel Dou [44] has investigated this question for two very
different black hole geometries, one in equilibrium (the 4
dimensional Schwarzschild metric) and one very far from
equilibrium (the conical horizon that represents the earliest
portion of a black hole formed from the collapse of a spherical
shell of matter).  For the Schwarzschild case, he made an ad hoc
approximation that reduced the problem to 2 dimensional
Schwarzschild and found, for a certain definition of ``near
horizon link'', that the number $N$ of such links has an
expectation value which reduces in the $\hbar\to{}0$ limit to
$c(\pi2/6)A$, where $A$ is the horizon area and $c$ is a
constant arising in the dimensional reduction.
(By $\hbar\to{0}$ I mean equivalently $l2/R2\to{0}$ 
%
%
where $l$ is the fundamental causet scale and $R$ the horizon
radius.)  For the expanding horizon case (again dimensionally
reduced from 4 to 2) he obtained exactly the same answer,
$c(\pi2/6)A$, despite the very different geometries.  Not only
is this a nontrivial coincidence, it represents the first time,
to my knowledge, that something like a number of horizon states
has been evaluated for any black hole far from equilibrium.  The
first step in solidifying and extending these results would be
to control the dimensional reduction from $4$ to $2$, evaluating
in particular the presently unknown coefficient $c$.  
Second, one should check that both null and spacelike hypersurfaces
$\Sigma$ yield the same results.
(The null case is the best studied to date.  Conceptually, it is
important for possible proofs of the generalized second law [45].)
Also one should assess the sensitivity of the answer
to changes in the definition of ``near horizon link'', since
there exist examples showing that the wrong definition can lead
to an answer of either zero or infinity.  And of course one
should extend the results to other black hole geometries beyond
the two studied so far.

\def\fns{$^\ddagger$}
Here is one of the definitions of ``near horizon link'' investigated by Djamel:
Let $H$ be the horizon of the black hole and $\Sigma$, as above, the
hypersurface for which we seek the entropy $S$.
The counting is meant to
yield the black hole contribution to $S$, corresponding to the section
$H\intersect\Sigma$ of the horizon.  We count pairs of sprinkled points
$(x,y)$ such that 
\item{}(i) $x\prec\Sigma, H$ and $y \succ \Sigma, H$.
\item{}(ii) $x \prec y$ is a link\footnote{\fns}{see glossary}
\item{}(iii) $x$ is maximal{\fns} in $(\past\Sigma)$ 
and $y$ is minimal{\fns} in $(\future\Sigma)\intersect(\future H)$.

\section{What are the ``observables'' of quantum gravity?}
Just as in the continuum the demand of diffeomorphism-invariance makes
it harder to formulate meaningful statements,\footnote{$^\dagger$}
{Think, for example, of the statement that light slows down when passing
 near the sun.}
so also for causets the demand of discrete general covariance has the
same consequence, bringing with it the risk that, even if we succeeded
in characterizing the covariant questions in abstract formal terms, we
might never know what they meant in a physically useful way.  I believe
that a similar issue will arise in every approach to quantum gravity,
discrete or continuous (unless of course general covariance is
renounced).\footnote{$^\flat$}
{In the context of canonical quantum gravity, this issue is called ``the
 problem of time''.  There, covariance means commuting with the
 constraints, and the problem is how to interpret quantities which do so
 in any recognizable spacetime language.  For an attempt in string
 theory to grapple with similar issues see [46].}

Within the context of the classical growth models described above, this
problem has been largely solved [47], the ``observables'' being
generated by ``stem-predicates''.  (`stem' is defined in the glossary).

\noindent {\it Added note:} The conjecture in [47] has been settled
in the affirmative by [48].

\section{How the large numbers of cosmology might be understood: a
``Tolman-Boltz\-mann'' cosmology}
Typical large number is ratio $r$ of diameter of universe to wavelength
of CMB radiation.  Idea is cycling of universe renormalizes
[49] coupling constants such that $r$ automatically gets big
after many bounces (no fine tuning).  Large numbers thus reflect large
age of universe.  See [50] and [51].

\section{Fields on a background causet}
See [52], [53], [54].

\section{Topology change}
See [55].

\section{Acknowledgments}
This research was partly supported by NSF grant PHY-0098488, by
a grant from the Office of Research and Computing of Syracuse
University, and by an EPSRC Senior Fellowship at Queen Mary
College, University of London.  I would like to express my warm
gratitude to Goodenough College, London for providing a splendid
living and working environment which greatly facilitated the
preparation of these notes.

\ReferencesBegin

\ref
[1]
R.D.~Sorkin, 
``A Specimen of Theory Construction from Quantum Gravity'',  
  in J.~Leplin (ed.),
  {\it The Creation of Ideas in Physics: Studies for a Methodology 
       of Theory Construction}
  (Proceedings of the Thirteenth Annual Symposium in Philosophy, 
       held Greensboro, North Carolina, March, 1989) 
  pp. 167-179
  (Number 55 in the University of Western Ontario Series in Philosophy of
            Science)
  (Kluwer Academic Publishers, Dordrecht, 1995)
  \eprint{gr-qc/9511063}


\ref
[2]
L.~Bombelli, J.~Lee, D.~Meyer and R.D.~Sorkin, ``Spacetime as a causal set'', 
  \journaldata {Phys. Rev. Lett.} {59} {521-524} {1987};
\refbreak
C.~Moore, ``Comment on `Space-time as a causal set','' 
  \journaldata{Phys. Rev. Lett.} {60} {655}{1988};
\refbreak
L.~Bombelli, J.~Lee, D.~Meyer and R.D.~Sorkin,  
  ``Bombelli et al. Reply'', 
 \journaldata {Phys. Rev. Lett.}{60}{656}{1988}.

\ref
[3]
L.~Bombelli, 
{\it Space-time as a Causal Set}, Ph.D. thesis, Syracuse University (1987)

\ref
[4]
G.~'t~Hooft, 
``Quantum gravity: a fundamental problem and some radical ideas'',
     in {\it Recent Developments in Gravitation }  
       (Proceedings of the 1978 Cargese Summer Institute) 
          edited by M. Levy and S. Deser
             (Plenum, 1979)

\ref
[5]
J.~Myrheim, ``Statistical geometry,'' CERN preprint TH-2538 (1978)

\ref
[6]
G.~Brightwell and R.~Gregory, 
``The Structure of Random Discrete Spacetime'',
\journaldata{Phys. Rev. Lett.}{66}{260-263}{1991}  

\ref
[7]
David D.~Reid, ``Discrete Quantum Gravity and Causal Sets'',
\journaldata{Canadian Journal of Physics}{79}{1-16}{2001}
\eprint{gr-qc/9909075}

\ref
[8]
Mituo Taketani, ``On formation of the Newton Mechanics'',
\journaldata{Suppl. Prog. Theor. Phys.}{50}{53-64}{1971}

\ref
[9]
Alfred Arthur Robb, {\it Geometry of Time and Space}
(Cambridge University Press, 1936)
(a revised version of {\it A theory of Time and Space} (C.U.P. 1914))

\ref
[10]
Hans Reichenbach:
\refbreak
\journaldata{Physikal. Zeitschr.}{22}{683}{1921};
and
\refbreak
{\it Axiomatik der relativistische Raum-Zeit-Lehre},
 translated into English as
 {\it Axiomatization of the theory of relativity}
 (Berkeley, University of California Press, 1969).

\ref
[11]
E.C. Zeeman, ``Causality Implies the Lorentz Group'', 
\journaldata{J.~Math. Phys.}{5}{490-493}{1964} 
%

\ref
[12]
Roger Penrose, {\it Techniques of Differential Topology in Relativity}, 
 AMS Colloquium Publications (SIAM, Philadelphia, 1972)

\ref
[13]
S.W. Hawking, A.R. King and P.J. McCarthy,
``A New Topology for Curved Space-Time which Incorporates
 the Causal, Differential and Conformal Structures'',
 \journaldata{J. Math. Phys.}{17} {174} {1976}

\ref
[14]
David Malament, ``The class of continuous timelike curves determines the
 toplogy of space-time'', 
\journaldata{J. Math. Phys}{18}{1399}{1977}

\ref
[15]
Bernhard Riemann, ``{\"U}ber die Hypothesen, welche der Geometrie zu Grunde
 liegen'',
  in 
 {\it The Collected Works of B. Riemann} (Dover NY 1953)

\ref
[16]
A.~Einstein, Letter to Walter D{\"a}llenbach, Nov. 1916, Item 9-072 
  translated and cited by Stachel in the following article, page 379.

\ref
[17]
A. Einstein, Letter to H.S. Joachim, August 14, 1954, Item 13-453 
  cited in J. Stachel,
  ``Einstein and the Quantum: Fifty Years of Struggle'', 
     in {\it From Quarks to Quasars, Philosophical Problems of Modern Physics},
        edited by R.G. Colodny
          (U. Pittsburgh Press, 1986), pages 380-381.

\ref
[18]
David Finkelstein, ``The space-time code'', 
  \journaldata{ Phys. Rev.} { 184}{1261} {1969}

\ref
[19]
 R.D.~Sorkin,
 ``Lisp Functions for Posets'',
   http://www.physics.syr.edu/$\widetilde{\phantom{n}}$sorkin 
   (version 1.0 1997, version 2.0 2002).
 \refbreak
Another poset package (using Maple) is [56].

\ref
[20]
Alan Daughton,			
 ``An investigation of the symmetric case of when causal sets can embed
  into manifolds'' 
 \journaldata{Class. Quant. Grav.}{15}{3427-3434}{1998}

\ref
[21]
G.~Brightwell and P.~Winkler, ``Sphere Orders'', 
\journaldata{Order} {6}{235-240} {1989}

\ref
[22]
Graham Brightwell, ``Models of Random Partial Orders'',   
 in {\it Surveys in Combinatorics, 1993},
 London Math. Soc. Lecture Notes Series {\bf 187}:~53-83,
 ed. Keith Walker
 (Cambridge Univ. Press 1993)

\ref
[23]
B. Pittel and R. Tungol, ``A Phase Transition Phenomenon in a Random Directed 
 Acyclic Graph'', 
(Ohio State preprint, 1998)
(submitted to {\it Combinatorics, Probability and Computing})

\ref
[24]
Klaus Simon, Davide Crippa and Fabian Collenberg,
``On the Distribution of the Transitive Closure in a Random Acyclic Digraph'',
{\it Lecture Notes in Computer Science} {\bf 726:} 345-356 (1993)

\ref
[25]
Charles M. Newman,
``Chain Lengths in Certain Random Directed Graphs'',
 {\it Random Structures and Algorithms} {\bf 3:} 243-253 (1992)

\ref
[26]
Jeong Han Kim and Boris Pittel,
``On tail distribution of interpost distance''
 (preprint, Ohio State University and Microsoft Corp., 1999)

\ref
[27]
Ioannis Raptis,  
``Algebraic Quantization of Causal Sets'' 
\journaldata{Int. J. Theor. Phys.}{39}{1233-1240}{2000} 
\eprint{gr-qc/9906103}

\ref
[28]
B.~Bollobas and G.~Brightwell,     
``The Height of a Random Partial Order: Concentration of Measure'',
 {\it Annals of Applied Probability} {\bf 2:} 1009-1018 (1992)

\ref
[29]
D.A.~Meyer: \refbreak
``Spherical containment and the Minkowski dimension of partial orders'', 
 \journaldata {Order} {10} {227-237} {1993};
and
\refbreak
{\it The Dimension of Causal Sets}, Ph.D. thesis, M.I.T. (1988).


\ref
[30]
R.D.~Sorkin, 
``On the Entropy of the Vacuum Outside a Horizon'',
  in B. Bertotti, F. de Felice and A. Pascolini (eds.),
  {\it Tenth International Conference on General Relativity and Gravitation
  (held Padova, 4-9 July, 1983), Contributed Papers}, 
  vol. II, pp. 734-736
  (Roma, Consiglio Nazionale Delle Ricerche, 1983);
  %
\refbreak
G.~'t~Hooft, ``On the quantum structure of a black hole'', 
   \journaldata{ Nuclear Phys. B} { 256}{727-745}{1985}

\ref
[31]  
L.~Bombelli and D.A.~Meyer, ``The origin of Lorentzian geometry,''
 \journaldata {Physics Lett. A} {141} {226-228} {1989}

\ref
[32] 
Johan Noldus,
``A new topology on the space of Lorentzian metrics on a fixed manifold''
\journaldata{Class. Quant. Grav}{19}{6075-6107}{2002}

\ref
[33]
Luca Bombelli, ``Statistical Lorentzian geometry and the closeness of
 Lorentzian manifolds'', 
\journaldata{J. Math. Phys.}{41}{6944-6958}{2000} 
\eprint{gr-qc/0002053}

\ref
[34]   
 R.D.~Sorkin: \refbreak 
``On the Role of Time in the Sum-over-histories Framework for Gravity'',
    paper presented to the conference on 
    The History of Modern Gauge Theories, 
    held Logan, Utah, July 1987, 
    published in \journaldata{Int. J. Theor. Phys.}{33}{523-534}{1994};
    and
\refbreak
 ``A Modified Sum-Over-Histories for Gravity'',
   reported in
  {\it 
   Proceedings of the International Conference on Gravitation and Cosmology, 
   Goa, India, 14-19 December, 1987},
   edited by 
   B.~R. Iyer, Ajit Kembhavi, Jayant~V. Narlikar, and C.~V. Vishveshwara,
   see pages 184-186 in the article by 
   D.~Brill and L.~Smolin: 
   ``Workshop on quantum gravity and new directions'', pp 183-191 
   (Cambridge University Press, Cambridge, 1988)
%
%


\ref
[35]     
R.D.~Sorkin,
``Forks in the Road, on the Way to Quantum Gravity'', talk 
   given at the conference entitled ``Directions in General Relativity'',
   held at College Park, Maryland, May, 1993,
   published in
   \journaldata{Int. J. Th. Phys.}{36}{2759--2781}{1997}   
   \eprint{gr-qc/9706002}
   %

\ref
[36]    
R.D.~Sorkin,
``Quantum Mechanics as Quantum Measure Theory'',
   \journaldata {Mod. Phys. Lett.~A} {9 {\rm (No.~33)}} {3119-3127} {1994}
   \eprint{gr-qc/9401003}

\ref
[37]
J.B.~Hartle, ``Spacetime Quantum Mechanics and the Quantum 
 Mechanics of Spacetime'',
 in B.~Julia and J.~Zinn-Justin (eds.),
 {\it Les Houches, session LVII, 1992, Gravitation and Quantizations}
 (Elsevier Science B.V. 1995)

\ref
[38]
David P.~Rideout and Rafael D.~Sorkin,
``A Classical Sequential Growth Dynamics for Causal Sets'',
 \journaldata{Phys. Rev. D}{61}{024002}{2000}
 \eprint{gr-qc/9904062}

\ref
[39]
David P. Rideout, {\it Dynamics of Causal Sets},
Ph.D. thesis (Syracuse University 2001)

\ref
[40]
T.~Jacobson,
``Black Hole Entropy and Induced Gravity'',
\eprint{gr-qc/9404039};
\refbreak
S.L.~Adler, ``Einstein gravity as a symmetry-breaking effect in quantum
   field theory'', 
   \journaldata{Reviews of Modern Physics} {54}{729-766} {1982};
\refbreak
V.A.~Kazakov,
 ``The appearance of matter fields from quantum fluctuations of 2D-gravity'',
 \journaldata{Mod. Phys. Lett. A}{4}{2125-2139}{1989} .

\ref
[41]
R.D.~Sorkin: \refbreak
``First Steps with Causal Sets'', 
  in R. Cianci, R. de Ritis, M. Francaviglia, G. Marmo, C. Rubano, 
     P. Scudellaro (eds.), 
  {\it General Relativity and Gravitational Physics} 
   (Proceedings of the Ninth Italian Conference of the same name, 
     held Capri, Italy, September, 1990), pp. 68-90
  (World Scientific, Singapore, 1991); and
\refbreak
``Spacetime and Causal Sets'', 
     in J.C. D'Olivo, E. Nahmad-Achar, M. Rosenbaum, M.P. Ryan, 
              L.F. Urrutia and F. Zertuche (eds.), 
    {\it Relativity and Gravitation:  Classical and Quantum} 
    (Proceedings of the {\it SILARG VII Conference}, 
      held Cocoyoc, Mexico, December, 1990), 
    pages 150-173
    (World Scientific, Singapore, 1991).

\ref
[42]      
R.D.~Sorkin, Two Talks given at the 1997 Santa Fe workshop:
``A Review of the Causal Set Approach to Quantum Gravity'' and
``A Growth Dynamics for Causal Sets'',
 presented at:
  ``New Directions in Simplicial Quantum Gravity''
  July 28 - August 8, 1997.
  The scanned in transparencies may be viewed at \linebreak
  http://t8web.lanl.gov/people/emil/Slides/sf97talks.html ;
\refbreak
Y.~Jack Ng and H.~van Dam, ``A small but nonzero cosmological constant'',
(Int. J. Mod. Phys D., to appear)
\eprint{hep-th/9911102}

\ref
[43]
Maqbool Ahmed, Scott Dodelson, Patrick Greene and Rafael D.~Sorkin,
``Everpresent $\Lambda$'',
\eprint{astro-ph/0209274}

\ref
[44]     
Djamel Dou,			
 ``Causal Sets, a Possible Interpretation for the Black Hole
 Entropy, and Related Topics'', 
 Ph.~D. thesis (SISSA, Trieste, 1999)
 \eprint{gr-qc/0106024}.
\hfil\break
See also Dj.~Dou and R.D.~Sorkin, ``{Black Hole Entropy as
Causal Links}'', {\it Foundations of Physics} (2003, to appear).

\ref
[45]      
 R.D.~Sorkin, 
``Toward an Explanation of Entropy Increase 
  in the Presence of Quantum Black Holes'',
  \journaldata {Phys. Rev. Lett.} {56} {1885-1888} {1986}

\ref
[46]
Edward Witten, ``Quantum Gravity in De Sitter Space''
\eprint{hep-th/0106109}

\ref
[47]
{Graham Brightwell}, {H. Fay Dowker}, {Raquel S. Garc{\'\i}a}, {Joe Henson} 
 and {Rafael D.~Sork\-in},
``General Covariance and the `Problem of Time' in a Discrete Cosmology'',
 in K.G.~Bowden, Ed., 	
 {\it Correlations}, Proceedings of the ANPA 23 conference,
 held August 16-21, 2001, Cambridge, England 
 (Alternative Natural Philosophy Association, London, 2002), pp 1-17
\eprint{gr-qc/0202097}

\ref
[48]
Graham Brightwell, Fay Dowker, Raquel Garcia, Joe Henson and Rafael D.~Sorkin,
``{}``Observables'' in Causal Set Cosmology'',
\eprint{gr-qc/0210061}

\ref
[49]
Xavier Martin, Denjoe O'Connor, David Rideout and Rafael D.~Sorkin,
``On the ``renormalization'' transformations induced by
  cycles of expansion and contraction in causal set cosmology'',
 \journaldata {Phys. Rev. D}{63}{084026}{2001}
 \eprint {gr-qc/0009063}

\ref
[50]
Rafael D.~Sorkin,
``Indications of causal set cosmology'',
 \journaldata {Int. J. Theor. Ph.} {39 (7)} {1731-1736} {2000}
 (an issue devoted to the proceedings of the Peyresq IV conference,
  held June-July 1999, Peyresq France) 
\eprint{gr-qc/0003043}

\ref
[51]   
Avner Ash and Patrick McDonald,
``Moment problems and the causal set approach to quantum gravity'',
\eprint{gr-qc/0209020}

\ref
[52]
A.R.~Daughton,			
 {\it The Recovery of Locality for Causal Sets and Related Topics},
  Ph.D. thesis, Syracuse University (1993)   

\ref
[53]
Roberto Salgado, Ph.D. thesis, Syracuse University (in preparation)

\ref
[54]
Richard F. Blute, Ivan T. Ivanov, Prakash Panangaden,
``Discrete Quantum Causal Dynamics'',
\eprint{gr-qc/0109053}


\ref
[55]
Fay Dowker,                                 
``Topology Change in Quantum Gravity'',
\eprint{gr-qc/0206020}

\ref
[56]
 John R. Stembridge (jrs@math.lsa.umich.edu),
 {\it A Maple Package for Posets: Version 2.1},
 (Ann Arbor 1998-may-10)
 available at www.math.lsa.umich.edu/$\widetilde{\phantom{n}}$jrs

\ref
[57]
David P.~Rideout and Rafael D.~Sorkin,
``Evidence for a continuum limit in causal set dynamics'',
 \journaldata {Phys. Rev. D}{63}{104011}{2001}
 \eprint{gr-qc/0003117}

\ref
[58]      
F.~Markopoulou and L.~Smolin, ``Causal evolution of spin networks'',
 \journaldata{ Nuc. Phys.} { B508} {409-430}{1997}
 \eprint{gr-qc/9702025}

\ref
[59]
R.D.~Sorkin, 
``Percolation, Causal Sets and Renormalizability'',
unpublished talk delivered at the RG2000 conference, held
in Taxco, Mexico, January 1999

\ref
[60]
David P. Rideout and Rafael D. Sorkin,
``Evidence for a Scaling Behavior in Causal Set Dynamics'',
(in preparation)

\ref
[61] 
L.~Smolin,  ``The fate of black hole singularities and the parameters
          of the standard models of particle physics and cosmology'',
 \eprint{gr-qc/9404011}

\ref
[62]
Roberto B.~Salgado, ``Some Identities for the Quantum Measure and its 
 Generalizations'',
 \journaldata{ Mod. Phys. Lett.} { A17}{711-728}{2002}
 \eprint{gr-qc/9903015}

\ref
[63]
Fay Dowker and Rafael D. Sorkin, 
``Toward an Intrinsic Definition of Relativistic Causality''
 (in preparation)

\ref
[64]
A.~Criscuolo and H.~Waelbroeck, ``Causal Set Dynamics: A Toy Model'', 
\journaldata{ Class. Quant. Grav.}{ 16}{1817-1832}{1999}
\eprint{gr-qc/9811088}

\ref
[65]
Joohan Lee and Rafael D. Sorkin,
``Limits of Sequential Growth Dynamics''
(in preparation)

\ref
[66]
T.~Jacobson,
``Black-hole evaporation and ultrashort distances'',
\journaldata{ Phys.~Rev. D}{ 44}{1731-1739}{1991)}

\ref
[67]
Curtis Greene et al.,
{\it A Mathematica Package for Studying Posets}
(Haverford College, 1990, 1994)
available at www.haverford.edu/math/cgreene.html

\ref
[68]
C.D.~Broad, {\it Scientific Thought}
(Harcourt Brace and Company, 1923)

\vfill\break


{
 \singlespace					   
 \vskip 0.5truein
 \centerline           {\bf GLOSSARY}
 \par\nobreak
 \noindent
 \parindent=2pt
 \parskip=6pt			
 }

\noindent {\it Major deviations from these definitions are rare in the
 literature but minor ones are common.  (In this glossary, we use the
 symbol $<$ rather than $\prec$.)}

\def\term{\leftskip=0truecm\rightskip=0truecm} 

\def\means{
   \par
   \leftskip=1.0truecm\rightskip=1.0truecm     
   \parindent=0pt}

\medskip

\term ancestor/descendant 
\means If $x< y$ then $x$ is an ancestor of $y$ and $y$ is a descendant of $x$.

\term antichain 
\means a trivial order in which no element is related to any other
       (cf. `chain')

\term causet = causal set = locally finite order

\term chain = linear order 
\means
    an order, any two of whose elements  are related.
    In particular, any linearly ordered subset of an order is a chain.
    $n$-chain = chain of $n$ elements

\term child/parent 
\means
    If $x < y$ is a link we can say that $x$ is a parent of $y$ and $y$
    a child of $x$. 
    One also says that ``$y$ covers $x$''. 

\term comparable
\means see `related'

\term covering relation, covers
\means see `child', `link'

\term descendant 
\means see `ancestor'

\term down-set = downward-set = past-set = order-ideal = ancestral set 
\means  a subset of an order that contains all the ancestors of its members

\term full stem 
\means
    A partial stem whose complement is the (exclusive) future of its top layer

\term future
\means see `past'

\term inf = greatest lower bound (cf. `sup')

\term interval
\means see `order-interval'


\term level 
\means
    In a past-finite causet the level of an element $x$ is the number of
    links in the longest chain $a < b < ... < c < x$.  Thus, level $0$
    comprises the minimal elements, level 1 is level 0 of the remainder,
    etc. 

\term linear extension
\means Let $S$ be a set and $<$ an order-relation on $S$.  A linear
    extension of $<$ is a second order-relation $\prec$ which extends
    $<$ and makes $S$ into a chain.

\term link = covering relation 
\means
    An irreducible relation of an order, that is, one not implied by the
    other relations via transitivity. 
    Of course we exclude pairs $(x x)$ from being links, in the case where
    such pairs are admitted into the order relation at all.

    (There's no inconsistency here with the notion that a ``chain'' ought
    to be made up of ``links'': the links in a chain are indeed links
    relative to the chain itself, even if they aren't links relative
    to the enveloping order.)

\term locally finite
\means  An order is locally finite iff all its order-intervals are finite.
        (cf. `past-finite')

\term maximal/minimal
\means A maximal/minimal element of an order is one without
    descendants/ancestors. 

\term natural labeling 
\means
    A natural labeling of a past-finite order is an assignment to its
    elements of labels 0 1 2 ... such that $x < y\implies label(x) < label(y)$.
    Thus it is essentially the same thing as a locally finite linear
    extension. 

\term order-interval (or just plain interval) 
\means
   The interval determined by two elements $a$ and $b$ 
   is the set, $\interval(a,b)=\braces{x|a<x<b}$.

\term order = ordered set = poset = partially ordered set = partial order
\means
    An order is a set of elements carrying a notion of ``ancestry'' or
    ``precedence''.  Perhaps the simplest way to express this concept
    axiomatically is to define an order as a transitive, irreflexive
    relation $<$ .   Many other, equivalent definitions are possible.
    In particular, many authors use the reflexive convention, in effect
    taking $\le$ as the defining relation.

    It is convenient to admit the empty set as a poset.

\term order-isomorphic
\means    isomorphic as posets 

\term origin = minimum element 
\means  a single element which is the ancestor of all others

\term originary 
\means   a poset possessing an origin is originary

\term parent
\means see `child'

\term partial stem (or just plain stem) 
\means a past set of finite cardinality 

\term partial post 
\means
An element $x$ of which no descendant has an ancestor spacelike to $x$.
The idea is $x$ is the progenitor of a ``child universe''
%
%

\term partially ordered set
\means see `order'

\term past/future 
\means $\past(x) = \braces{y | y < x}$ ,  $\future(x) = \braces{y | x < y}$ 

\term past-finite 
\means  An order is past-finite iff all its down-sets are finite.
        (cf. `locally finite')

\term path = saturated chain 
\means
    a chain all of whose links are also links of the enveloping poset
    (saturated means it might be ``extended'' but it can't be ``filled in'')

\term poset
\means see `order'

\term post 
\means
    An element such that every other element is either its ancestor or
    its descendant: a one-element slice.

\term preorder = preposet = acyclic relation = acyclic digraph 
\means A preorder is a relation whose transitive closure is an order.

\term pseudo-order = transitive relation (possibly with cycles)


\term related = comparable 
\means
    Two elements $x$ and $y$ are `related' (or `comparable') if 
    $x < y$ or $y < x$.

\term slice  = maximal antichain 
\means (where maximal means it can't be enlarged and remain an antichain)

    equivalently, every $x$ in the causet is either in the slice or
    comparable to one of its elements. 

    equivalently, its inclusive past is a full stem

\term spacelike = incomparable 
\means
   Two elements $x$ and $y$ are spacelike to each other
   ($x\,\natural\,y$) iff they are unrelated (ie neither 
   $x < y$ nor $y < x$).

\term stem
\means see `partial stem'

\term sup = least upper bound  (cf. `inf')

\term transitively reduced 
\means
    The ``transitive reduction'' of an order is its Hasse digraph,
    an acyclic relation containing only links.

\end


(prog1    'now-outlining
  (Outline 
      "
     "......"
      "
   "\\\\message" 
   "\\\\section"
   "\\\\appendi"
   "\\\\Referen"	
   "\\[69" 	
      "
   "\\\\subsectio"
   "\\\\term....."	
   ))